# MC-GenRef: Annotation-Free Mammography Microcalcification Segmentation with Generative Posterior Refinement

Hyunwoo Cho, *Member, IEEE*,  Yeeun Kwon, *Student Member, IEEE*, Min Jung Kim, and Yangmo Yoo, *Member, IEEE*

**Abstract**—Microcalcification (MC) analysis is clinically important in screening mammography because clustered puncta can be an early sign of malignancy, yet dense MC segmentation remains challenging: targets are extremely small and sparse, dense pixel-level labels are expensive and ambiguous, and cross-site shift often induces texture-driven false positives and missed puncta in dense tissue. We propose MC-GenRef, a real dense-label-free framework that combines high-fidelity synthetic supervision with test-time generative posterior refinement (TT-GPR). During training, real negative mammogram patches are used as backgrounds, and physically plausible MC patterns are injected through a lightweight image formation model with local contrast modulation and blur, yielding exact image-mask pairs without real dense annotation. Using only these synthetic labeled pairs, MC-GenRef trains a base segmentor and a seed-conditioned rectified-flow (RF) generator that serves as a controllable generative prior. During inference, TT-GPR treats segmentation as approximate posterior inference: it derives a sparse seed from the current prediction, forms seed-consistent RF projections, converts them into case-specific surrogate targets through the frozen segmentor, and iteratively refines the logits with overlap-consistent and edge-aware regularization. On INbreast, the synthetic-only initializer achieved the best Dice ($0.80 \pm 0.17$) without real dense annotations, while TT-GPR improved miss-sensitive performance to Recall $0.89 \pm 0.14$ and FNR $0.11 \pm 0.14$, with strong class-balanced behavior (Bal.Acc. $0.95 \pm 0.07$, G-Mean $0.94 \pm 0.10$). On an external private Yonsei cohort ($n = 50$), TT-GPR consistently improved the synthetic-only initializer under cross-site shift, increasing Dice and Recall while reducing FNR. These results suggest that test-time generative posterior refinement is a practical route to reduce MC misses and improve robustness without additional real dense labeling.

*Index Terms*— Microcalcification segmentation, Screening mammography, Synthetic supervision, Test-time adaptation, Generative posterior refinement

## I. INTRODUCTION

MICROCALCIFICATIONS (MCs) are among the earliest radiographic signs of breast malignancy and a frequent trigger for recall in screening mammography [1]. They can precede the appearance of a visible mass, and clinical assessment depends not only on their presence but also on morphology and distribution, such as scattered puncta, clustered patterns, heterogeneous density, and linear or segmental arrangement [2, 3]. This motivates dense mapping, including segmentation or dense probability estimation, because it supports cluster characterization, targeted review, longitudinal comparison, and integration into downstream CAD logic [4, 5]. At the same time, MC segmentation is an unusually adversarial vision problem: targets are tiny, often only a few pixels; positives are extremely sparse; and the background is dominated by structured, high-frequency texture from glandular tissue, acquisition noise, and vendor-specific processing [5, 6]. In this regime, the confound is systemic: a bright dot can be a true calcification, a textured gland edge, a compression artifact, or a local noise fluctuation, and these modes overlap heavily in appearance.

These properties pose significant challenges for both classical pipelines and modern deep models [6-9]. Detecting subtle MCs requires sensitivity to weak high-frequency cues, yet mammograms contain abundant high-frequency structures that are not specific to malignant pathology. Increasing sensitivity therefore tends to inflate texture-driven false positives (FPs), reducing trust and limiting clinical utility. A practical goal is thus to provide stable, morphology-consistent, lesion-specific evidence that remains reliable across tissue density, background complexity, and acquisition variability. Prior deep learning studies have addressed these objectives, but

Date of submission: 6 April 2026.

We thank Seung Yeop Yang, BPH, for assistance with data collection and data management during the establishment and annotation of the database. The private microcalcification cohort was collected under Institutional Review Board approval (IRB No. 4-2026-0043) at Severance Hospital, Yonsei University.

This work was supported by the Korea Medical Device Development Fund grant funded by the Korea government (the Ministry of Science and ICT, the Ministry of Trade, Industry and Energy, the Ministry of Health & Welfare, Republic of Korea, the Ministry of Food and Drug Safety) (KMDF202011A01-03, KMDF202011A01-04).

Hyunwoo Cho and Yeeun Kwon are with the Department of Electronic Engineering, Sogang University, Seoul 04107, Korea (e-mails: hwcho.research@gmail.com; ky1511@sogang.ac.kr)

Min Jung Kim is with the Department of Radiology, Research Institute of Radiological Science, and Center for Clinical Imaging Data Science (CCIDS), Yonsei University College of Medicine, Seoul 03722, Korea. (e-mail: MINES@yuhs.ac)

Yangmo Yoo is with the Department of Electronic Engineering, Sogang University, Seoul 04107, Korea and with the Department of Biomedical Engineering, Sogang University, Seoul 04107, Korea. (e-mail: ymyoo@sogang.ac.kr) (Corresponding author) most rely on standard full supervision [6-9]. This reliance is a bottleneck: dense pixel-level MC masks are expensive and often ambiguous, and the ambiguity scales with target size [10]. When lesions span only a few pixels, boundary decisions become subjective and sensitive to windowing, sampling, and reader preference, amplifying inter-reader variability. Although curated datasets exist [11, 12], they still provide limited supervision for multi-site and multi-vendor settings, and this limitation can hinder cross-site generalization.

Weak supervision, such as image labels, boxes, or points, reduces labeling burden, but it often lacks the spatial precision needed for tiny-object segmentation and can struggle to suppress texture-driven false positives in dense tissue [13]. Hand-crafted approaches, including multi-scale filters [14], difference-of-Gaussian or Hessian cues [15], blob rules, and morphology heuristics, are interpretable and label-light, yet they are brittle under vendor variability and can degrade when structured background dominates. Fully supervised deep segmentation can outperform classical methods on matched distributions, but it may still rely on high-frequency shortcuts and remain sensitive to training-set characteristics and domain shift.

A pragmatic way to bypass real dense labels is synthetic supervision, where dense masks are free by construction.

Similar to prior studies on synthetic tumors in computed tomography images [16], MC synthesis is particularly natural because MC-like patterns can be injected into mammogram patches while retaining exact pixel masks. This can bootstrap a segmentor to learn canonical MC morphology and substantially scale the training volume. However, synthetic-only training often induces a large domain gap to real data [17]. Models may exploit subtle injection artifacts or oversimplified intensity priors. More fundamentally, the dominant confound, namely case-specific structured background, does not follow a simple generative rule. As a result, synthetic training alone often yields either missed subtle MCs that deviate from the synthetic distribution or FP blobs that latch onto dense texture.

In an annotation-free setting, synthetic fidelity therefore becomes pivotal. If synthetic MCs are rendered as toy blobs with unrealistic contrast, scale, blur, or cluster geometry, the learned decision boundary will not transfer well to real mammograms. Higher-fidelity synthesis can instead provide a semantic anchor at zero labeling cost by respecting physical scale, appearance variability, local contrast behavior, and acquisition blur. Still, synthetic supervision alone cannot fully capture the physical complexity of real MC appearance or the case-specific structured background that dominates false alarms. To bridge this gap, we propose MC-GenRef, an annotation-free

**(A) Synthetic training phase**  **(B) Test-time refinement phase**

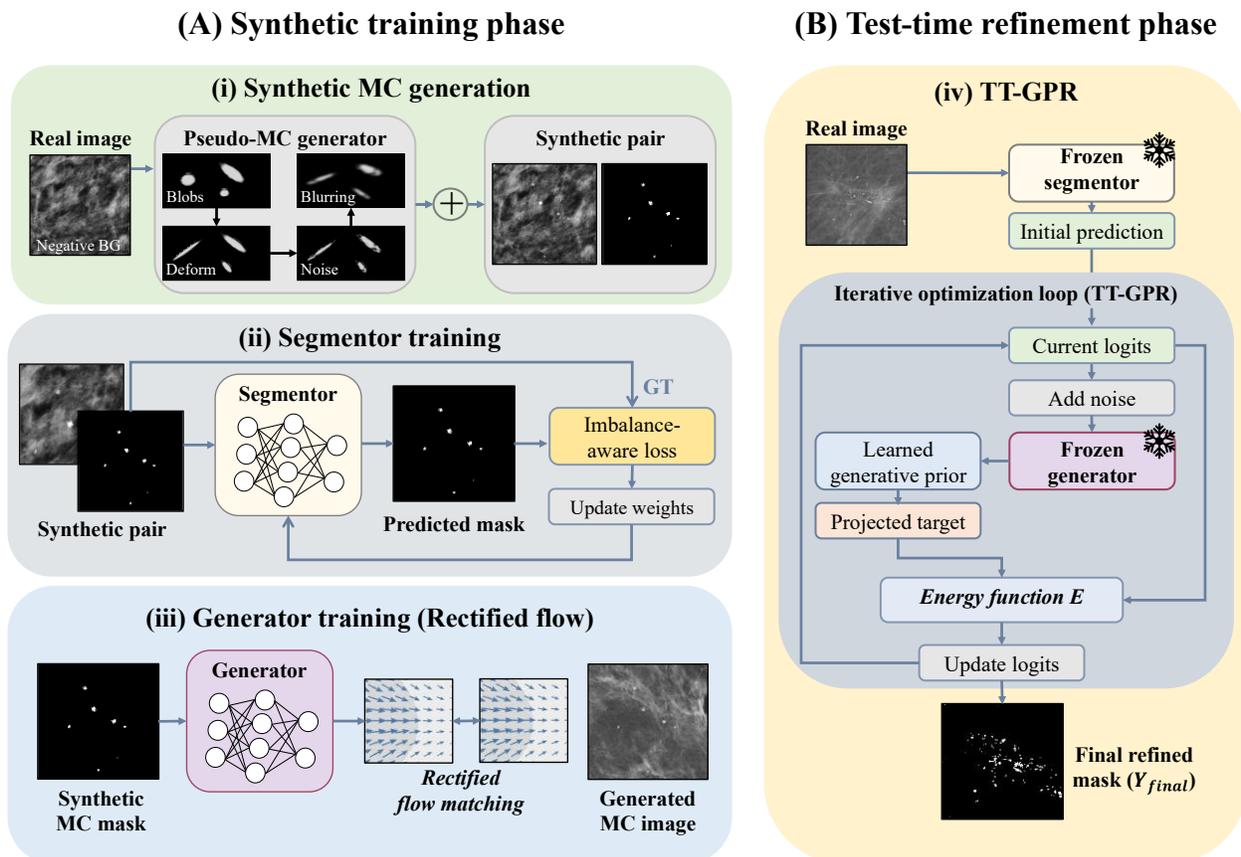

**Fig. 1.** Overview of MC-GenRef. The framework consists of synthetic MC generation on real mammographic backgrounds, synthetic-only training of a segmentor and a seed-conditioned RF generator, and test-time generative posterior refinement on real mammograms.

framework based on generative posterior refinement. We first construct a simple yet effective synthetic pipeline in which real mammogram patches serve as backgrounds to preserve authentic texture statistics, and MC patterns are injected via a lightweight lesion model and an image-formation module that captures blur and local contrast, producing paired images and masks consistent with mammographic appearance. While this synthetic process provides a plausible MC manifold for both the discriminative segmentor and the generative prior, the remaining synthetic-to-real mismatch is handled at inference. Rather than treating segmentation as a one-shot forward pass, we cast it as posterior inference. Given an image x, we seek a mask that is both plausible under a learned conditional generative prior and consistent with the observed image under structured background. Generative models offer a direct way to encode such priors [18, 19], and their projections can provide case-specific surrogate supervision that enables refinement without dense real annotations.

MC-GenRef is built around test-time generative posterior refinement for MC segmentation. It uses two learned components trained only on synthetic labeled pairs: a base segmentor (Seg) and a conditional rectified flow (RF) generative prior (Prior). At inference on real mammograms, we run TT-GPR (Test-time Generative Posterior Refinement), an iterative procedure that alternates between (i) forming RF-based projection targets by anchoring the current probability map and applying the Prior, and (ii) updating Seg logits by minimizing a compact energy that promotes prior-consistent, prevalence-aware, and spatially regular solutions. This design suppresses pepper-noise FPs while preserving tiny objects. TT-GPR is plug-and-play: it operates directly on logits and uses a separate Prior, enabling refinement of different segmentors without architectural changes. Crucially, MC-GenRef is dense-label-free: neither Seg nor Prior uses real pixel-level MC masks for training. High-fidelity synthetic data provides the semantic and morphological anchor, while test-time posterior refinement provides case-adaptive correction for common failures, particularly texture-driven FPs and fragmented tiny positives. In our experiments, MC-GenRef achieved sensitivity competitive with, and in some cases higher than, supervised baselines while requiring no dense real annotations. We also collected an external dataset for cross-site validation, where MC-GenRef improved consistently while supervised baselines often degraded under domain shift. Our code will be available at https://github.com/hwcho-research/MC-GenRef.

## II. Materials and Methods

MC-GenRef targets microcalcification (MC) segmentation under a real dense-label-free training setting, that is, without using real pixel-level MC masks for model training. The framework consists of three components: high-fidelity synthetic MC generation on real mammographic backgrounds, synthetic-only training of a base segmentor and a seed-conditioned rectified-flow (RF) generator, and test-time generative posterior refinement (TT-GPR) on real mammograms. The segmentor provides an initial dense prediction, while the RF generator provides a controllable generative prior that is used only at inference to refine the prediction in a case-adaptive manner. Figure 1 summarizes the overall pipeline.

### A. Problem setup and synthetic MC generation

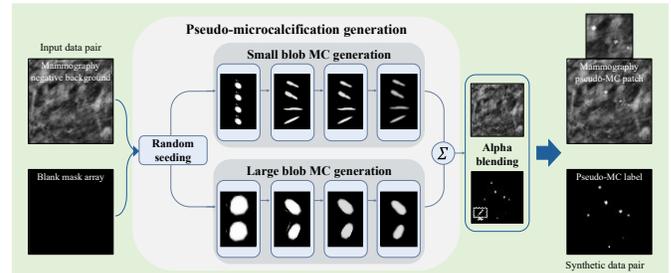

**Fig. 2** Synthetic MC generation on real mammographic backgrounds. Real negative mammogram patches are used as backgrounds, and synthetic MC patterns are injected with realistic size, contrast, clustering, and blur, yielding label-accurate image-mask pairs.

Let $x \in [-1,1]^{H \times W}$ denote a normalized mammogram patch and $y \in [0,1]^{H \times W}$ the corresponding MC mask. The base segmentor $S_\theta$ outputs a logit map $l$ and a probability map $p$ :

$$l = S_\theta(x), p = \sigma(l). \qquad (1)$$

Because real dense MC masks are expensive and ambiguous, we construct a synthetic labeled set $\mathcal{D}_{\text{syn}} = \{(x_s, y_s)\}$ and reserve real mammograms for test-time refinement and evaluation. Synthetic training starts from real negative mammogram patches sampled as backgrounds, so that the spatial spectrum and structured high-frequency appearance of mammographic tissue are preserved. This is important because the dominant error mode in MC segmentation is not random noise alone, but confusion with case-specific background texture.

On top of each real negative background $x_b$, we generate synthetic MC patterns with variable number, size, blur, intensity, and spatial arrangement. Cluster centers are sampled stochastically, and the individual puncta are distributed either compactly around a cluster center or along short line-like structures to mimic clinically relevant clustered or segmental patterns. Each calcification is modeled as a small compact support whose physical size is mapped into pixels, and a soft MC intensity field is formed by summing Gaussian-like kernels with variable amplitude and spread. To approximate acquisition blur and post-processing, the soft field is convolved with a blur kernel and then injected into the real background with local contrast modulation:

$$\Delta x = h * a_s, \quad x_s = \text{clip}(x_b + \kappa \odot \Delta x, -1,1), \qquad (2)$$

where $a_s$ is the synthetic MC field, $h$ is a blur kernel, and $\kappa$ is a local contrast scaling factor derived from background statistics. The binary support of the injected puncta defines the exact mask $y_s$. This formulation preserves realistic mammographic texture while providing label-accurate supervision.

Given a synthetic pair $(x_s, y_s)$, the segmenter prediction is $p_s = \sigma(S_\theta(x_s))$. The segmenter is trained using a combination of Dice and focal-Tversky losses [20]:

$$\mathcal{L}_{seg} = \lambda_D \mathcal{L}_{Dice}(p_s, y_s) + \lambda_T \mathcal{L}_{FT}(p_s, y_s). \quad (3)$$

Here, the Tversky weighting is chosen to penalize false negatives more strongly than false positives, so that the initializer remains sensitive to subtle MCs under extreme sparsity. This is intentional because later test-time refinement will suppress implausible activations using the learned prior.

### B. Seed-conditioned rectified-flow prior

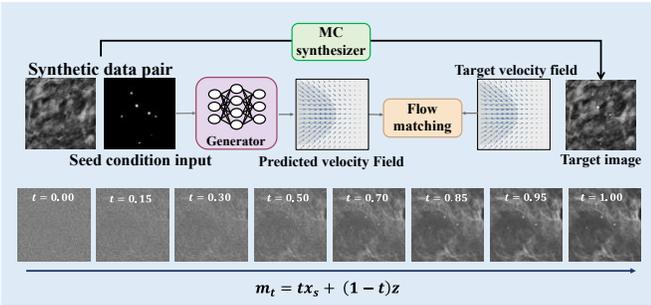

**Fig. 3** Seed-conditioned rectified-flow generator. The RF model learns to generate an MC-consistent mammogram patch from noise conditioned on a sparse seed mask and is later used as a projection operator during test-time refinement.

Synthetic supervision alone provides a useful morphological anchor, but it does not fully bridge the gap to real mammograms. To provide a controllable prior for inference-time correction, we train a seed conditioned RF generator $G_\phi$ in image space. For each synthetic pair, a sparse seed mask $s_s = \mathcal{A}(y_s)$ is extracted from the full MC mask, for example by point sampling or thinning. This seed acts as a compact condition that preserves lesion location and coarse structure while avoiding direct use of the full mask at inference.

With Gaussian noise $z \sim \mathcal{N}(0, I)$ and time $t \sim \mathcal{U}[0, 1]$, rectified flow defines an interpolated state

$$m_t = tx_s + (1-t)z. \quad (4)$$

The generator predicts a conditional velocity field $G_\phi(m_t, t; s_s)$, and training is performed by flow matching:

$$\mathcal{L}_{RF} = \mathbb{E}_{x_s, y_s, z, t}\left[\left\|G_\phi(tx_s + (1-t)z, t; s_s) - (x_s - z)\right\|_2^2\right]. \quad (5)$$

After training, $G_\phi$ provides a practical projection operator $\Pi_\phi(x; s, t, z)$ that maps an observed patch $x$, a sparse seed $s$, and random noise $z$ to a seed-consistent image sample. Importantly, the RF generator is not used as a standalone detector. Instead, it generates case-specific image projections that are converted into surrogate teacher targets through the frozen segmentor during test-time refinement.

### C. Test-time generative posterior refinement

At inference, the ground-truth mask is unavailable. We therefore refine the segmentor output directly on the real test image. Starting from the initial logits

$$l^{(0)} = S_\theta(x), \quad (6)$$

TT-GPR iteratively constructs case-specific correction targets from the current prediction. At iteration $i$, the current probability map is $p^{(i)} = \sigma(l^{(i)})$. A sparse seed $s^{(i)} = \mathcal{A}(p^{(i)})$ is extracted from the current estimate, and a projection time is chosen using a coarse-to-fine schedule:

$$t_i = t_{start} + \frac{i}{\max(1, I-1)}(t_{end} - t_{start}). \quad (7)$$

Early iterations therefore allow stronger prior-driven correction, whereas later iterations remain closer to the observed image.

Using a fresh Gaussian sample $z^{(i)}$, the RF generator produces a seed-consistent projected image

$$\hat{x}^{(i)} = \Pi_\phi(x; s^{(i)}, t_i, z^{(i)}), \quad (8)$$

which is then passed through the frozen segmentor to obtain a surrogate teacher target:

$$q^{(i)} = \sigma\left(S_\theta(\hat{x}^{(i)})\right). \quad (9)$$

During refinement, $q^{(i)}$ is treated as a fixed target, that is, gradients are not propagated through the projection step or the frozen segmentor. This stabilizes optimization while preserving the role of the generator as a case-adaptive prior.

To compare the current prediction $p = \sigma(l)$ with the projected target $q = q^{(i)}$, we define the soft Tversky statistics

$$TP = \sum_u p_u q_u,$$
$$FP = \sum_u p_u(1 - q_u), \quad FN = \sum_u (1 - p_u)q_u, \quad (10)$$

and the corresponding Tversky index

$$TI(p, q) = \frac{TP + \varepsilon}{TP + \alpha_T FP + \beta FN + \varepsilon} \quad (11)$$

$$\mathcal{L}_{TI}(p, q) = 1 - TI(p, q). \quad (12)$$

Here, $\alpha_T$ is fixed and $\beta$ controls the false-negative penalty. A larger $\beta$ encourages miss reduction, which is particularly important in MC screening.

The refinement energy at iteration $i$ is then defined as

$$E^{(i)}(l; x) = w_T \mathcal{L}_{TI}(\sigma(l), q^{(i)}) + w_S \|\sigma(l) - q^{(i)}\|_2^2 + w_E \mathcal{L}_{edge}(\sigma(l); x). \quad (13)$$

The first term encourages agreement with the RF-based projection target, the second stabilizes the stochastic target, and the third aligns the prediction with image edges to suppress pepper-noise false positives while preserving tiny structures.

Logits are updated by gradient descent:

$$\tilde{l}^{(i+1)} = l^{(i)} - \eta \nabla_l E^{(i)}(l^{(i)}; x), \quad (14)$$

followed by a small edge-guided high-pass mixing step in logit space,

$$l^{(i+1)} = \tilde{l}^{(i+1)} + \rho e(x), \quad (15)$$

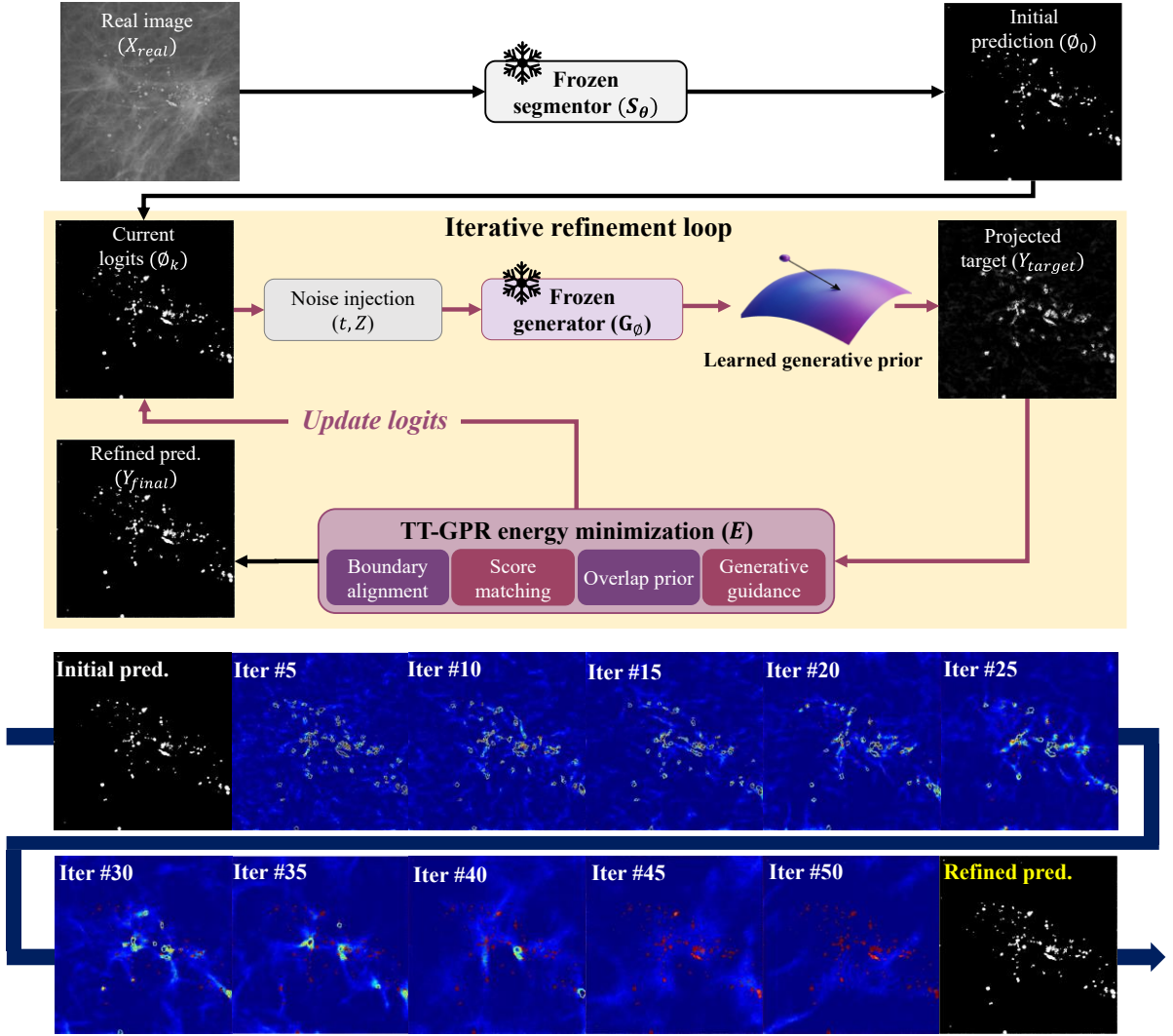

**Fig. 4** Test-time generative posterior refinement (TT-GPR). At inference, TT-GPR iteratively derives a seed from the current prediction, forms an RF-based projection target, and updates the logits to improve sensitivity while suppressing pepper-noise false positives.

where $e(x)$ is the normalized image edge map, $\eta$ is the refinement learning rate, and $\rho$ is the high-pass logit mixing rate. After $I$ iterations, the final refined output is

$$p^{\star} = \sigma\big(l^{(I)}\big). \tag{16}$$

This procedure can be interpreted as approximate posterior inference. The RF generator provides a seed-conditioned generative prior, and the frozen segmentor translates projected samples into case specific supervisory signals. In this way, TT-GPR moves the prediction toward solutions that are both image-consistent and prior-consistent, rather than relying on a single forward pass.

### D. Implementation details and evaluation

Both the segmentor and the RF generator adopt a Restormer backbone [21]. The RF generator is trained with AdamW [22] using a learning rate of $2 \times 10^{-4}$ for 300 k steps, and exponential moving average weights are used for stable sampling and projection [23]. The segmentor is then trained on synthetic labeled pairs with the same optimizer. TT-GPR hyperparameters are selected with Optuna [24], and the final settings are summarized in Table 1. Synthetic training is built on real negative mammogram patches sampled from the RSNA screening mammography dataset [25]. A small labeled subset of INbreast [12] is used only for validation and hyperparameter selection, while the model training itself remains real dense-label-free. This setup follows the original experimental design while keeping the method centered on synthetic-only learning and test-time refinement.

We evaluate the proposed method on the INbreast validation cohort and on an external Yonsei cohort ($n = 50$) collected under institutional review board approval and annotated by an experienced breast imaging radiologist. The Yonsei cohort is used strictly for evaluation and provides a realistic cross-site test under acquisition and texture shift. Comparison methods include CSD [9], HDoGReg-FPN [8], HDoGReg-Hybrid [8], DeepMiCa [7], a fully supervised baseline trained on INbreast,

| Symbol | Meaning | Value |
|--------|---------|-------|
| $I$ | Number of refinement iterations | 50 |
| $t_{start}$ | Start time of coarse-to-fine schedule | 0.09053149415704413 |
| $t_{end}$ | End time of coarse-to-fine schedule | 0.8052230726911125 |
| $w_T$ | Weight of Tversky-style overlap term | 128.15533098670937 |
| $w_S$ | Weight of MSE stabilization term | 57.88546958560159 |
| $w_E$ | Weight of edge-alignment term | 220.70931647468294 |
| $\beta$ | False-negative penalty in Tversky index | 2.6140537072408576 |
| $\eta$ | Refinement learning rate | 0.4945043784053065 |
| $\rho$ | High-pass logit mixing rate | 2.8422788048947973 |

the synthetic-only initializer, and the full MC-GenRef with TT-GPR.

We evaluate MC segmentation as pixel-wise binary classification with confusion-matrix counts $TP, FP, FN$, and $TN$. Overlap fidelity is measured by the Dice coefficient,

$$\text{Dice} = \frac{2TP}{2TP + FP + FN} \qquad (17)$$

Sensitivity is measured by Recall,

$$\text{Recall} = \frac{TP}{TP + FN}, \qquad (18)$$

and the corresponding miss probability is the false negative rate,

$$\text{FNR} = \frac{FN}{FN + TP}. \qquad (19)$$

To account for severe class imbalance, we also report specificity,

$$\text{TNR} = \frac{TN}{TN + FP}, \qquad (20)$$

and balanced accuracy,

$$\text{Bal.Acc.} = \frac{\text{Recall} + \text{TNR}}{2}. \qquad (21)$$

We further summarize joint sensitivity and specificity using the geometric mean,

$$\text{G-Mean} = \sqrt{\text{Recall} \cdot \text{TNR}}. \qquad (22)$$

Finally, to emphasize recall while still accounting for false alarms, we report precision,

$$\text{Precision} = \frac{TP}{TP + FP}, \qquad (23)$$

and the $F_2$ score,

$$F_2 = \frac{5 \cdot \text{Precision} \cdot \text{Recall}}{4 \cdot \text{Precision} + \text{Recall}}. \qquad (24)$$

Together, these metrics capture overlap quality (Dice), missed-lesion risk (Recall and FNR), class-imbalance-robust discrimination (Balanced Accuracy and G-Mean), and recall-oriented detection performance ($F_2$).

## III. RESULTS

Figure 5 shows a representative INbreast case with a clustered microcalcification pattern embedded in heterogeneous fibroglandular texture, where the main challenge is recovering numerous tiny puncta while avoiding spurious activations on structured background. Classical baselines either under-activate and miss subtle puncta or produce sparse and unstable detections that do not form a coherent cluster. Deep baselines increase cluster coverage, but fragmentation and residual false positives remain in ambiguous high-frequency regions. Our synthetic-only initializer already recovers a substantial portion of the cluster, suggesting that high-fidelity synthetic supervision provides a useful semantic anchor even without real dense labels. After TT-GPR, the prediction becomes more spatially coherent within the cluster: fragmented responses consolidate into compact micro-blobs, cluster density increases, and off cluster pepper-noise responses are reduced. This qualitative pattern is consistent with case-adaptive posterior refinement rather than a simple global threshold shift.

Cohort-level INbreast box plots are summarized in Figure 6. Quantitatively, CSD (Wang) yielded Dice $0.14 \pm 0.05$, Recall $0.11 \pm 0.04$, FNR $0.89 \pm 0.04$, balanced accuracy $0.56 \pm 0.02$, G-Mean $0.33 \pm 0.07$, and $F_2$ $0.12 \pm 0.04$. HDoGReg (FPN) improved to Dice $0.47 \pm 0.24$, Recall $0.42 \pm 0.24$, FNR $0.58 \pm 0.24$, balanced accuracy $0.71 \pm 0.12$, G-Mean $0.60 \pm 0.23$, and $F_2$ $0.43 \pm 0.24$, while HDoGReg (Hybrid) achieved Dice $0.49 \pm 0.24$, Recall $0.36 \pm 0.20$, FNR $0.64 \pm 0.20$, balanced accuracy $0.68 \pm 0.10$, G-Mean $0.56 \pm 0.22$, and $F_2$ $0.40 \pm 0.22$. DeepMiCa showed substantially stronger performance with Dice $0.76 \pm 0.20$, Recall $0.73 \pm 0.19$, FNR $0.27 \pm 0.19$, balanced accuracy $0.87 \pm 0.10$, G-Mean $0.85 \pm 0.13$, and $F_2$ $0.73 \pm 0.21$. The fully supervised reference achieved Dice $0.75 \pm 0.14$, Recall $0.86 \pm 0.14$, FNR $0.14 \pm$ $0.14$, balanced accuracy $0.93 \pm 0.07$, G-Mean $0.92 \pm 0.08$, and $F_2$ $0.81 \pm 0.13$. Notably, our synthetic only initializer remained highly competitive without any real dense labels, achieving the highest Dice of $0.80 \pm 0.17$ together with Recall $0.76 \pm 0.20$, FNR $0.24 \pm 0.20$, balanced accuracy $0.88 \pm 0.10$, G-Mean $0.86 \pm 0.16$, and $F_2$ $0.78 \pm 0.18$. TT-GPR then shifted the operating point toward miss reduction, reaching Recall $0.89 \pm 0.14$, FNR $0.11 \pm 0.14$, balanced accuracy $0.95 \pm 0.07$, G-Mean $0.94 \pm 0.10$, Dice $0.72 \pm 0.18$, and $F_2$ $0.80 \pm 0.16$. These results indicate that the synthetic-only model provides the strongest overlap-oriented initializer, whereas TT-GPR improves miss-sensitive and class-balanced performance by reducing false negatives in this extremely sparse setting.

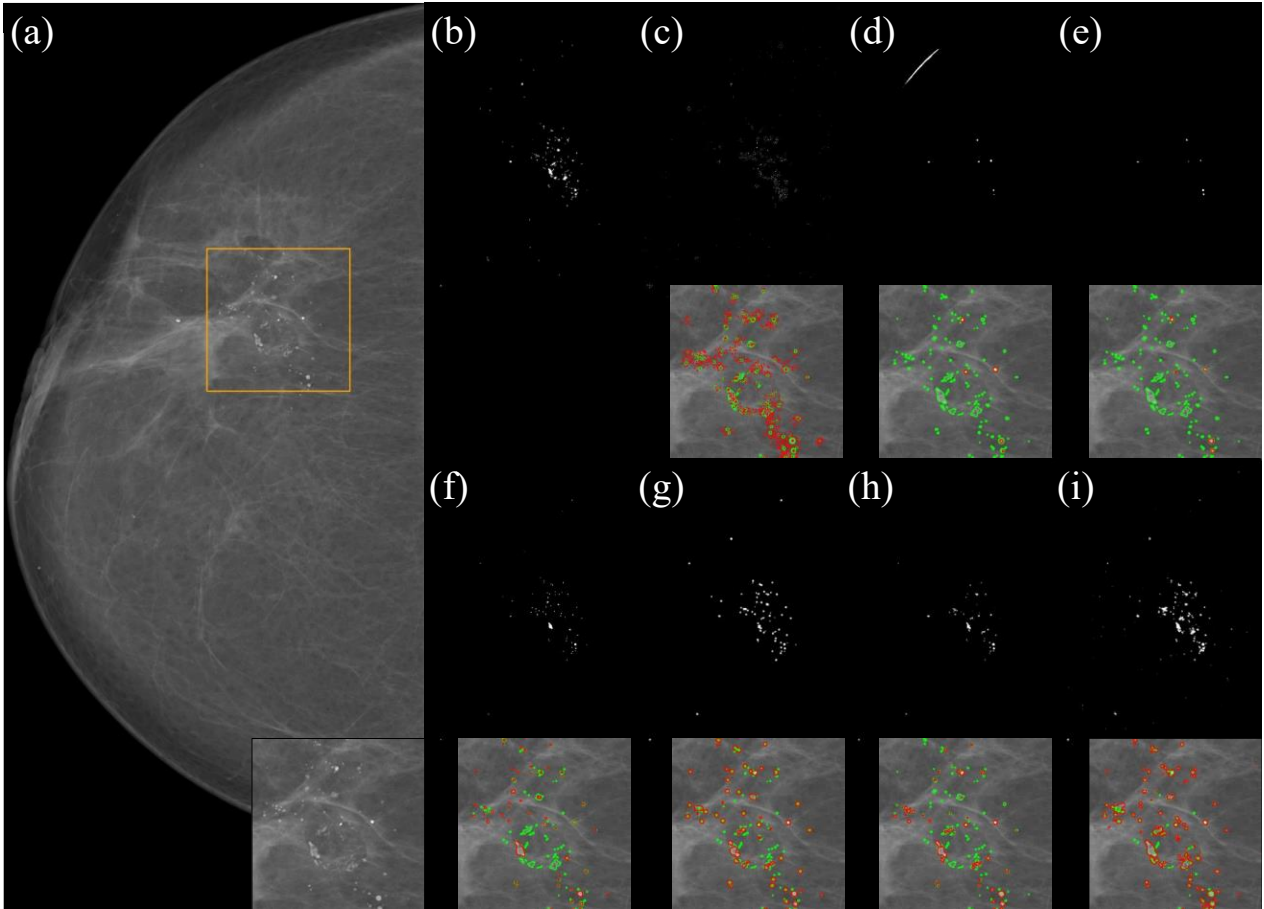

**Fig. 5** Qualitative comparison on INbreast (case 1): binary MC maps in the evaluated ROI. (a) Full mammogram with evaluated ROI (orange box). (b) Ground-truth MC mask in the ROI. (c) CSD (Wang). (d) HDoGReg (FPN). (e) HDoGReg (Hybrid). (f) DeepMiCa. (g) Supervised. (h) Ours (w/o TT-GPR; synthetic-only initializer). (i) Ours (TT-GPR refined). All methods are visualized in a common binary format to highlight tiny-object fragmentation and pepper-noise false positives.

Figure 7 presents a representative qualitative result on the external Yonsei cohort, where dense site-specific parenchymal texture makes tiny calcification puncta particularly difficult to distinguish from structured background. In this regime, CSD remains sparse and fragmented, while the HDoGReg variants often produce coarse blob-like responses around the lesion region. Although such responses may appear prominent, they do not localize punctate MCs faithfully. DeepMiCa and the supervised reference concentrate predictions more closely around the lesion, but still exhibit local misses or small off-cluster activations under challenging background contrast. In contrast, our synthetic-only initializer already forms a compact clustered hypothesis within the ROI, indicating that the learned MC morphology transfers meaningfully across sites without real dense-label training. TT-GPR further sharpens this estimate by strengthening weak puncta inside the cluster and suppressing isolated off-cluster specks, yielding a cleaner and more cluster-faithful binary map.

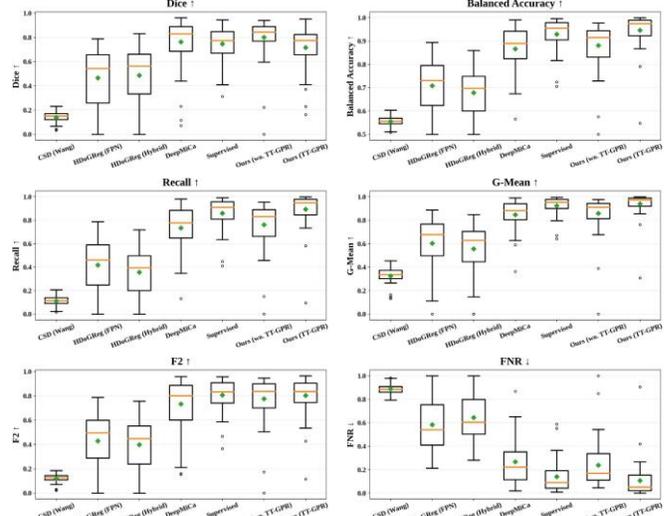

**Fig. 6** Cohort-level INbreast performance (per-case box plots) over six metrics. Methods: CSD (Wang), HDoGReg (FPN), HDoGReg (Hybrid), DeepMiCa, Supervised, Ours (w/o TT-GPR), and Ours (TT-GPR). Metrics: Dice, Balanced Accuracy, Recall, G-Mean, F2, and FNR (lower is better). Boxes indicate interquartile ranges with median lines; markers denote means

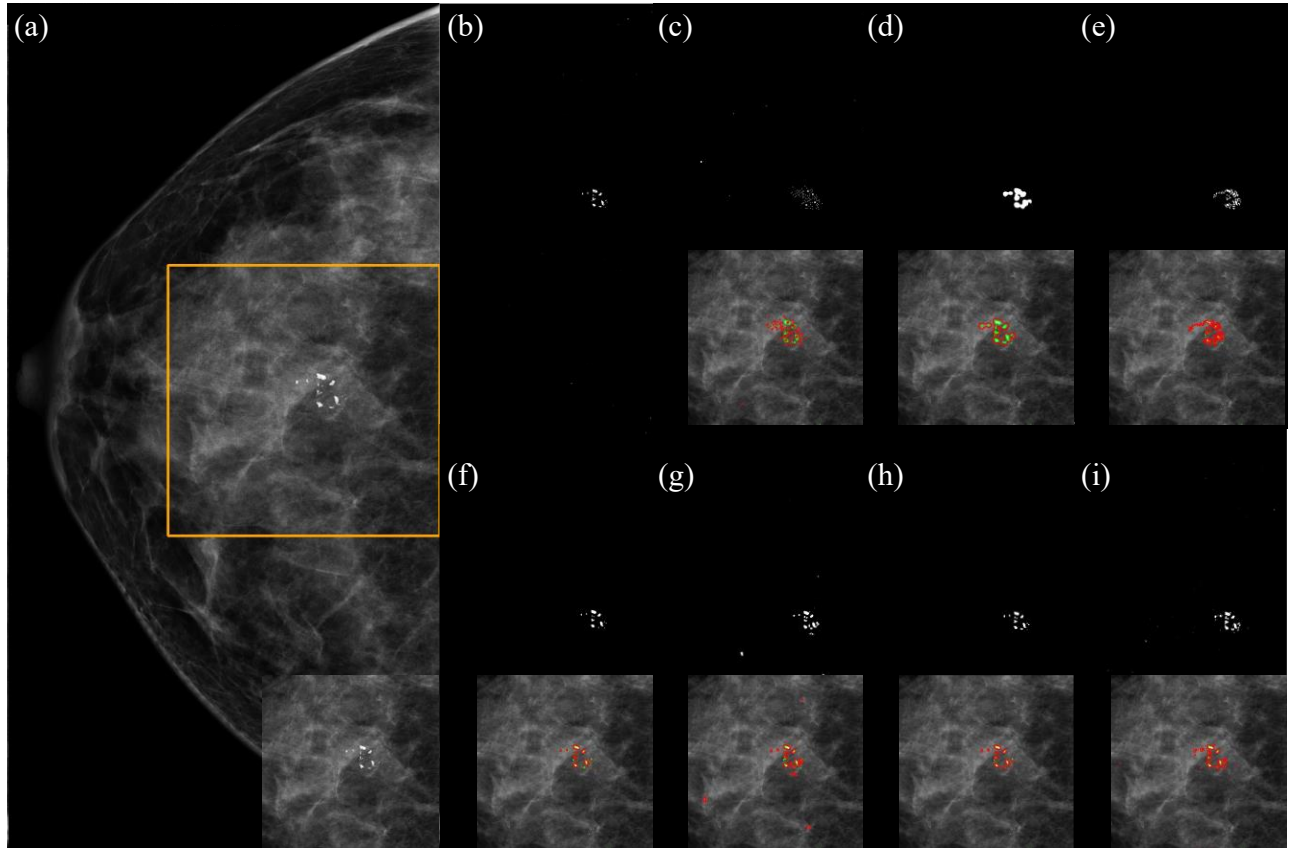

**Fig. 7** Qualitative results on the private Yonsei dataset (representative case). (a) Full-field mammogram with the evaluation ROI highlighted (yellow box). (b) Radiologist annotation (binary MC mask) within the ROI. (c–g) Competing methods: CSD (Wang), HDoGReg (FPN), HDoGReg (Hybrid), DeepMiCa, and the fully supervised reference. (h) Ours (w/o TT-GPR; synthetic-only initializer). (i) Ours (TT-GPR refined). Under dense background texture, HDoGReg variants often produce coarse, blob-like components that appear prominent but are poorly localized for punctate MCs, whereas TT-GPR sharpens the clustered hypothesis by suppressing isolated spurious responses while preserving cluster-consistent puncta

Cohort-level Yonsei box plots are summarized in Figure 8. Under cross-site domain shift, all methods degraded relative to INbreast, which is expected for such extremely sparse targets. Quantitatively, CSD (Wang) produced Dice $0.19 \pm 0.07$, Recall $0.20 \pm 0.09$, FNR $0.80 \pm 0.09$, balanced accuracy $0.60 \pm 0.04$, G-Mean $0.44 \pm 0.12$, and $F_2$ $0.19 \pm 0.07$. HDoGReg (FPN) reached the highest Recall among the compared baselines at $0.45 \pm 0.29$, but with very low Dice of $0.16 \pm 0.08$, together with FNR $0.55 \pm 0.29$, balanced accuracy $0.72 \pm 0.14$, G-Mean $0.62 \pm 0.26$, and $F_2$ $0.24 \pm 0.14$, indicating a sensitivity-heavy but poorly localized operating point. HDoGReg (Hybrid) achieved Dice $0.27 \pm 0.15$, Recall $0.31 \pm 0.22$, FNR $0.69 \pm 0.22$, balanced accuracy $0.66 \pm 0.11$, G-Mean $0.51 \pm 0.23$, and $F_2$ $0.29 \pm 0.18$. DeepMiCa transferred poorly, with Dice $0.19 \pm 0.20$, Recall $0.17 \pm 0.19$, FNR $0.83 \pm 0.19$, balanced accuracy $0.59 \pm 0.10$, G-Mean $0.34 \pm 0.24$, and $F_2$ $0.17 \pm 0.19$. The fully supervised reference achieved Dice $0.20 \pm 0.17$, Recall $0.38 \pm 0.26$, FNR $0.62 \pm 0.26$, balanced accuracy $0.69 \pm$

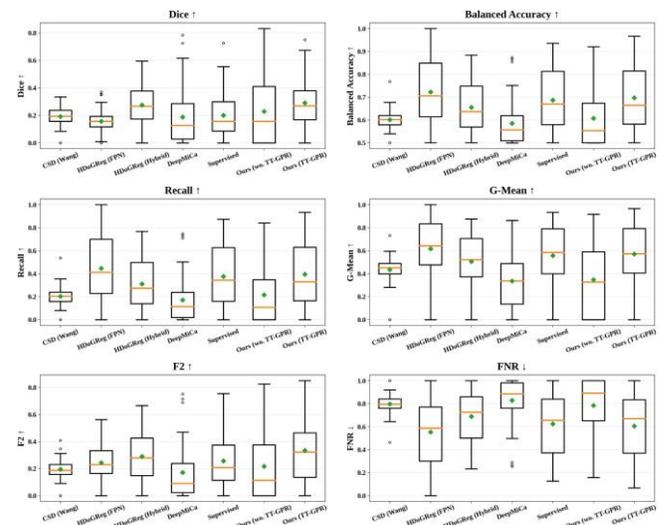

**Fig. 8** Cohort-level Yonsei dataset performance (per-case box plots) over six metrics. Methods: CSD (Wang), HDoGReg (FPN), HDoGReg (Hybrid), DeepMiCa, Supervised, Ours (w/o TT-GPR), and Ours (TT-GPR). Metrics: Dice, Balanced Accuracy, Recall, G-Mean, F2, and FNR (lower is better). Boxes indicate interquartile ranges with median lines; markers denote means

0.13, G-Mean $0.56 \pm 0.26$, and $F_2 0.26 \pm 0.20$. Our synthetic-only initializer yielded Dice $0.23 \pm 0.24$, Recall $0.22 \pm 0.26$, FNR $0.78 \pm 0.26$, balanced accuracy $0.61 \pm 0.13$, G-Mean $0.35 \pm 0.31$, and $F_2 0.22 \pm 0.25$. After TT-GPR, performance improved consistently to Dice $0.29 \pm 0.18$, Recall $0.39 \pm 0.29$, FNR $0.61 \pm 0.29$, balanced accuracy $0.70 \pm 0.14$, GMean $0.57 \pm 0.26$, and $F_2 0.33 \pm 0.22$. Notably, TT-GPR achieved the best Dice and the best $F_2$ on Yonsei, while ranking within the top two for the remaining metrics, supporting the view that test-time generative posterior refinement improves localizer faithful sensitivity under realistic cross-shift.

## IV. DISCUSSION

This study presents MC-GenRef, a dense-label-free framework that combines high-fidelity synthetic supervision with test-time generative posterior refinement for microcalcification segmentation. The central idea is to separate representation learning from case-adaptive correction. Synthetic training on real mammographic backgrounds provides a strong morphological anchor without requiring real pixel level MC masks, while TT-GPR refines the prediction at inference by enforcing consistency with a seed conditioned generative prior. Unlike a fixed threshold shift or conventional post-processing, the refinement is instance-adaptive and directly targets the dominant failure mode of MC analysis: missed puncta and texture-driven false positives under structured background.

On INBreast, the synthetic-only initializer already achieved the highest Dice among all compared methods, indicating that carefully designed synthetic supervision can capture sufficiently informative MC morphology even without real dense labels. TT-GPR then shifted the operating point toward clinically relevant miss reduction. Recall improved from $0.76 \pm 0.20$ to $0.89 \pm 0.14$ and FNR decreased from $0.24 \pm 0.20$ to $0.11 \pm 0.14$, while balanced accuracy and G-Mean also increased. Although Dice decreased from $0.80 \pm 0.17$ to $0.72 \pm 0.18$, this pattern suggests that TT-GPR is not simply enlarging the prediction. Rather, it reallocates the solution toward recovering tiny positives that are easily missed in one-shot inference, while maintaining strong class-balanced behavior.

The external Yonsei cohort provides a more stringent evaluation under cross-site shift. As expected, all methods degraded relative to INbreast, reflecting the difficulty of transferring tiny-object segmentation across differences in acquisition characteristics, tissue texture, and local contrast. In this setting, the fully supervised reference and several existing baselines transferred inconsistently. HDoGReg-FPN achieved relatively high recall, but often through coarse, blob-like responses that were poorly localized for punctate MCs. In contrast, TT-GPR consistently improved the synthetic-only initializer, increasing Dice from $0.23 \pm 0.24$ to $0.29 \pm 0.18$, Recall from $0.22 \pm 0.26$ to $0.39 \pm 0.29$, balanced accuracy from $0.61 \pm 0.13$ to $0.70 \pm 0.14$, G-Mean from $0.35 \pm 0.31$ to $0.57 \pm 0.26$, and F2 from $0.22 \pm 0.25$ to $0.33 \pm 0.22$ while reducing FNR from $0.78 \pm 0.26$ to $0.61 \pm 0.29$. Notably, the refined model achieved the best Dice and F2 on the external cohort, supporting the view that posterior refinement improves robustness beyond in-domain optimization.

These results also highlight an important evaluation perspective. For extremely sparse targets such as MCs, Dice alone does not fully represent clinical utility. A system that recovers more true puncta, even at the cost of some overlap redistribution, may be preferable in screening settings where missed lesions carry disproportionate risk. In this sense, the synthetic-only initializer and TT-GPR play complementary roles: the former provides a strong overlap-oriented starting point, and the latter adds a miss-reduction mechanism that adapts the final prediction to the local structure of each case.

Several limitations remain. First, TT-GPR adds test-time computation, which may be non-trivial in largescale screening workflows. Second, refinement depends on hyperparameters such as the number of steps, projection schedule, and regularization weights, and the best operating point may vary across sites. Third, the Yonsei cohort is relatively small and private, which limits statistical power and acquisition diversity. Finally, although pixel-wise metrics are appropriate for benchmarking segmentation accuracy, future work would be strengthened by lesion-level or cluster-level endpoints that more directly reflect clinical decision-making. Practical next steps therefore include faster refinement, automatic parameter selection, and broader multi-institutional validation.

Overall, MC-GenRef provides a practical bridge between synthetic supervision and real-world deployment for MC analysis. A high-fidelity synthetic initializer offers strong baseline performance without real dense labels, and test-time posterior refinement improves sensitivity and reduces misses under domain shift through case-adaptive correction. These findings support generative posterior refinement as a promising direction for label-efficient and robust microcalcification analysis.

## V. CONCLUSION

We presented MC-GenRef, a dense-label-free framework for microcalcification segmentation that couples high-fidelity synthetic supervision with test-time generative posterior refinement. By constructing label-accurate synthetic MC pairs on real mammographic backgrounds, the method learns a strong initializer without requiring real pixel-level MC masks. At inference, TT-GPR refines the prediction through seed-conditioned generative projections, improving robustness to structured background and cross-site shift. On INbreast, the synthetic-only initializer achieved the best Dice ($0.80 \pm 0.17$), while TT-GPR improved Recall to $0.89 \pm 0.14$ and reduced FNR to $0.11 \pm 0.14$ with strong classbalanced performance. On the external Yonsei cohort, TT-GPR consistently improved the synthetic-only initializer and achieved the best Dice ($0.29 \pm 0.18$) and F2 ($0.33 \pm 0.22$). These results suggest that test-time generative posterior refinement can complement synthetic supervision by reducing missed MCs without additional real dense labeling. Future work will focus on accelerating refinement and validating the framework across larger multi-institutional cohorts.

## REFERENCES

[1] R. Bonfiglio, M. Scimeca, N. Urbano *et al.*, "Breast microcalcifications: Biological and diagnostic perspectives," 30, Taylor & Francis, 2018, pp. 3097-3099.

[2] M. E. Anderson, M. S. Soo, R. C. Bentley *et al.*, "The detection of breast microcalcifications with medical ultrasound," *The Journal of the Acoustical Society of America,* vol. 101, no. 1, pp. 29-39, 1997.

[3] G. Tse, P. H. Tan, A. L. Pang *et al.*, "Calcification in breast lesions: pathologists' perspective," *Journal of clinical pathology,* vol. 61, no. 2, pp. 145-151, 2008.

[4] H. Du, M. M.-S. Yao, S. Liu *et al.*, "Automatic calcification morphology and distribution classification for breast mammograms with multi-task graph convolutional neural network," *IEEE Journal of Biomedical and Health Informatics,* vol. 27, no. 8, pp. 3782-3793, 2023.

[5] M. A. Gavrielides, J. Y. Lo, and C. E. Floyd Jr, "Parameter optimization of a computer-aided diagnosis scheme for the segmentation of microcalcification clusters in mammograms," *Medical Physics,* vol. 29, no. 4, pp. 475-483, 2002.

[6] G. Valvano, G. Santini, N. Martini *et al.*, "Convolutional neural networks for the segmentation of microcalcification in mammography imaging," *Journal of healthcare engineering,* vol. 2019, no. 1, pp. 9360941, 2019.

[7] A. Gerbasi, G. Clementi, F. Corsi *et al.*, "DeepMiCa: Automatic segmentation and classification of breast MIcroCAlcifications from mammograms," *Computer Methods and Programs in Biomedicine,* vol. 235, pp. 107483, 2023.

[8] C. Marasinou, B. Li, J. Paige *et al.*, "Segmentation of breast microcalcifications: A multi-scale approach," *arXiv preprint arXiv:2102.00754,* 2021.

[9] J. Wang, and Y. Yang, "A context-sensitive deep learning approach for microcalcification detection in mammograms," *Pattern recognition,* vol. 78, pp. 12-22, 2018.

[10] K. Wang, M. Hill, S. Knowles-Barley *et al.*, "Improving segmentation of breast arterial calcifications from digital mammography: good annotation is all you need." pp. 130-146.

[11] R. S. Lee, F. Gimenez, A. Hoogi *et al.*, "A curated mammography data set for use in computer-aided detection and diagnosis research," *Scientific data,* vol. 4, no. 1, pp. 170177, 2017.

[12] I. C. Moreira, I. Amaral, I. Domingues *et al.*, "Inbreast: toward a full-field digital mammographic database," *Academic radiology,* vol. 19, no. 2, pp. 236-248, 2012.

[13] Y. Zeng, Y. Zhuge, H. Lu *et al.*, "Multi-source weak supervision for saliency detection." pp. 6074-6083.

[14] H. Min, S. S. Chandra, N. Dhungel *et al.*, "Multi-scale mass segmentation for mammograms via cascaded random forests." pp. 113-117.

[15] B. Thangaraju, I. Vennila, and G. Chinnasamy, "Detection of microcalcification clusters using hessian matrix and foveal segmentation method on multiscale analysis in digital mammograms," *Journal of digital imaging,* vol. 25, no. 5, pp. 607-619, 2012.

[16] L. Wu, J. Zhuang, Y. Zhou *et al.*, "Large-scale generative tumor synthesis in computed tomography images for improving tumor recognition," *Nature Communications,* vol. 16, no. 1, pp. 11053, 2025.

[17] A. Ackaouy, N. Courty, E. Vallée *et al.*, "Unsupervised domain adaptation with optimal transport in multi-site segmentation of multiple sclerosis lesions from MRI data," *Frontiers in computational neuroscience,* vol. 14, pp. 19, 2020.

[18] H. Chung, and J. C. Ye, "Score-based diffusion models for accelerated MRI," *Medical image analysis,* vol. 80, pp. 102479, 2022.

[19] Y. Song, J. Sohl-Dickstein, D. P. Kingma *et al.*, "Score-based generative modeling through stochastic differential equations," *arXiv preprint arXiv:2011.13456,* 2020.

[20] N. Abraham, and N. M. Khan, "A novel focal tversky loss function with improved attention u-net for lesion segmentation." pp. 683-687.

[21] S. W. Zamir, A. Arora, S. Khan *et al.*, "Restormer: Efficient transformer for high-resolution image restoration." pp. 5728-5739.

[22] I. Loshchilov, and F. Hutter, "Decoupled weight decay regularization," *arXiv preprint arXiv:1711.05101,* 2017.

[23] D. Morales-Brotons, T. Vogels, and H. Hendrikx, "Exponential moving average of weights in deep learning: Dynamics and benefits," *arXiv preprint arXiv:2411.18704,* 2024.

[24] T. Akiba, S. Sano, T. Yanase *et al.*, "Optuna: A next-generation hyperparameter optimization framework." pp. 2623-2631.

[25] C. Carr, F. Kitamura, G. Partridge *et al.*, "RSNA screening mammography breast cancer detection," *Kaggle,* 2022.